\input{aipcheck}

\documentclass[
    ,final            % use final for the camera ready runs
  ]
  {aipproc}

\layoutstyle{6x9}
\begin{document}

\title{Detection of gravitational waves by pulsar timing}


\author{Alexander Rodin}{
  address={Russia, 142290, Moscow region, Pushchino, PRAO}
}

\begin{abstract}
A new approach to the problem of gravitational waves detection based on simultaneous timing of several pulsars and subsequent expansion of the post-fit timing data into components of different spectral kind (with different spectral indices) is proposed. Presence of a signal caused by stochastic gravitational waves in spectral components is tested with the two-point angular correlation function as proposed in the pulsar timing array. This new approach was applied to timing data of a few millisecond pulsars and allowed to detect a signature similar to one predicted for gravitational wave background at relatively high confidence level: correlation coefficient between experimental and theoretical two-point correlations $\rho=0.82\pm0.07$.
\end{abstract}

\maketitle

%%%%%%%%%%%%%%%%%%%%%%%%%%%%%%%%%%%%%%%%%%%%
%% MAINMATTER
%%%%%%%%%%%%%%%%%%%%%%%%%%%%%%%%%%%%%%%%%%%%

\section{Introduction}

Fractional instability of some pulsars is comparable with the one of atomic standards and reaches the level $10^{-15}$. Such a property allows to use pulsar timing for solution of various astronomical and metrological problems. In this paper a method of detection of the gravitational waves (GW) by pulsar timing array based on the two-point correlation function (eq.~\ref{two-point}) is further developed \cite{HellingsDowns:1983,jenet:2005,ZhaoZhang:2003}.
\begin{equation}\label{two-point}
\zeta(\theta)=\frac32x\log(x)-\frac x4+\frac12+\frac12\delta(x),
\end{equation}
where $x=(1-\cos\theta)/2$, $\theta$ is angular distance between pulsars, $\delta$ is delta-function.

Many radio observatories started pulsar timing observational programs with the final aim of detection of the gravitational waves. By the year 2005 the Kalyazin Radio Astronomical Observatory (KRAO) of the Astro Space Center of the Lebedev Physical Institute already had 8-9 year time series of pulsar timing data of 6 millisecond pulsars (PSRs J0613-0200, J1640+2224, J1643-1224, J1713+0747, J1939+2134, J2145-0750). After publication of the paper \cite{jenet:2005} method of the two-point correlation function was applied to the data (pulsar post-fit residuals of time of arrivals (TOAs)) and showed that it can not be used directly since TOA residuals always display a presence of so called timing noise that prevents detection of the signature (\ref{two-point}) clearly.

However the additive noise presented as a phase variations in all frequency standards (atomic clocks or pulsars) has distinct features which allows to separate one kind of the noise from another. Each kind of the additive noise can be described in form $h_n/\omega^{n}$, where $n$ is spectral index of the power spectrum of the phase variations, $\omega$ is Fourier frequency, $h_n$ is the amplitude of the noise. Each spectral index $n$ corresponds to noise caused by different physical phenomena \cite{blandford:1984}. E.g., $n=0$ corresponds to the measurement noise, $n=3$ corresponds to the phase variations due to interstellar medium \cite{blandford:1984}, $n=5$ corresponds to the variations caused by the primordial gravitational wave background (GWB) \cite{bertotti:1983}.

As an experience of frequency standards exploitation shows, noise with more steep spectrum begins to dominate at more long time interval since the amplitudes $h_n$ rapidly diminish with increasing of $n$. This property of the frequency standards can be used for separation of the additive noise into components with different spectral indices by an appropriate mathematical method. A possible presence of the GWB in the components can be detected with the two-point correlation function. 
The singular spectrum analysis (SSA) was used for expansion of pulsar timing data into components \cite{golyandina:2001}. As fig.~5 from the paper \cite{Rodin:2011} shows, the SSA-method indeed expands the input stochastic signal (TOA residuals) into components with different spectral indices. This property is particularly noticeable for low and high frequency components.

\section{Observations}

The pulsars  PSR J0613-0200, J1640+2224, J1643-1224, J1713+0747, J1939+2134 and J2145-0750 were observed with the fully steerable RT-64 radio telescope of KRAO \cite{Ilyasov:2004, Ilyasov:2005, Potapov:2003}. The AS-600 instrumental facility of the Pushchino Radio Astronomy Observatory (Astro Space Center, Lebedev Physical Institute) was used for the registration \cite{Oreshko:2000}. The pulsar pulses were accumulated by a spectrum analyser in two circular polarizations, 80 channels in each, with a channel frequency band 40 kHz. The observing sessions were conducted, on average, once every two weeks. The total signal integration time in each session was about 2 hours.

The topocentric TOAs were determined by fitting the session-summed pulsar pulse profile into a reference template with a hight signal to noise ratio. We computed the barycentric TOAs, determined the TOA residuals and refined the pulsar timing parameters by minimizing the residuals by the least-squares method using the software Tempo \cite{Taylor:1989}. The DD model \cite{DamourDeruelle:1986} was used to refine and compute the pulsar orbital parameters. The astrometric, spin and orbital parameters of all six pulsars can be found in the paper \cite{Rodin:2011}. Fig.~1 shows the post-fit TOA residuals of 6 pulsars.
%-------------------------------------------------------------
\begin{figure}\label{fig1}
\includegraphics[scale=0.7]{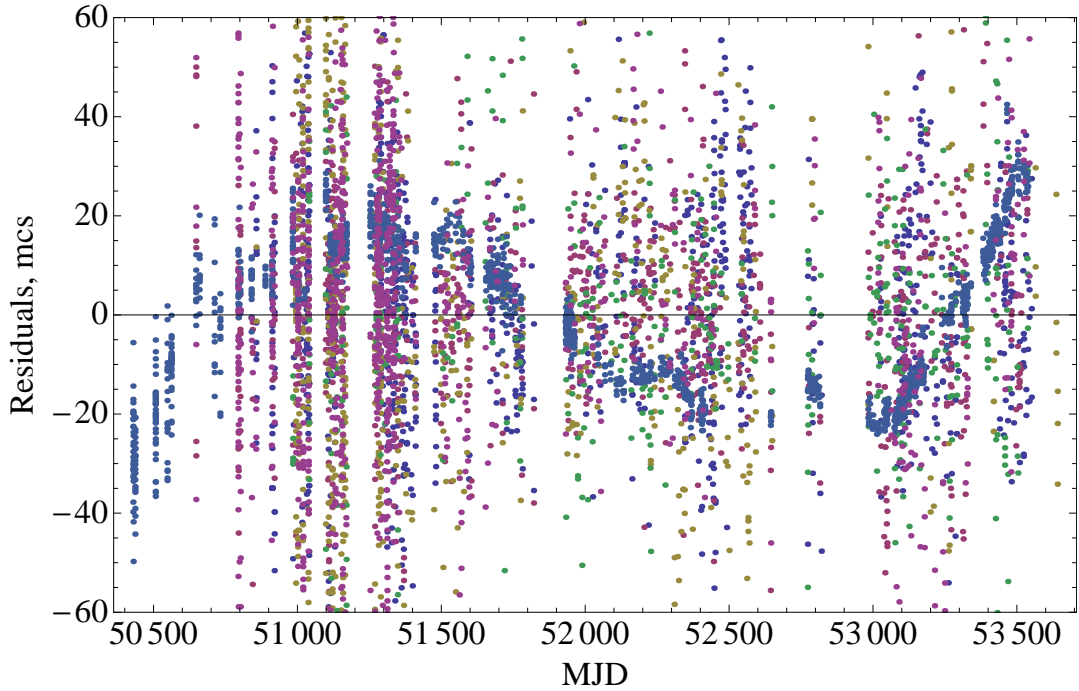}
\caption{Post-fit timing residuals of six millisecond pulsars observed at the Kalyazin Radio Astronomy Observatory.}
\end{figure}
%-------------------------------------------------------------

\section{Computer simulation}

For computer simulation the coordinates of above six pulsars were taken. A total of $N=50$ generations were made in accordance with the number of TOAs. The amplitude of the GW was set to constant 
$h=1$, right ascension of the GW was uniformly distributed in the range $0^h \leq \alpha<24^h$ and sine of declination of the GW was uniformly distributed in the range $-1<\sin \delta < 1$. Next, the angular coefficients were calculated on the basis of formula (1) from \cite{HellingsDowns:1983}. Fig.~2a shows results of the computer simulation. On the basis of the simulation one can conclude that with the $M=6$ pulsars and $N=50$ TOAs the two-point correlation function $\zeta(\theta)$ is detected with the amplitude signal-to-noise ratio $SNR\simeq 10$.
%-------------------------------------------------------------
\begin{figure}\label{fig2}
\begin{minipage}{0.5\linewidth}
\vbox{\hbox{\hspace*{4.4cm}a)}\includegraphics[scale=0.72]{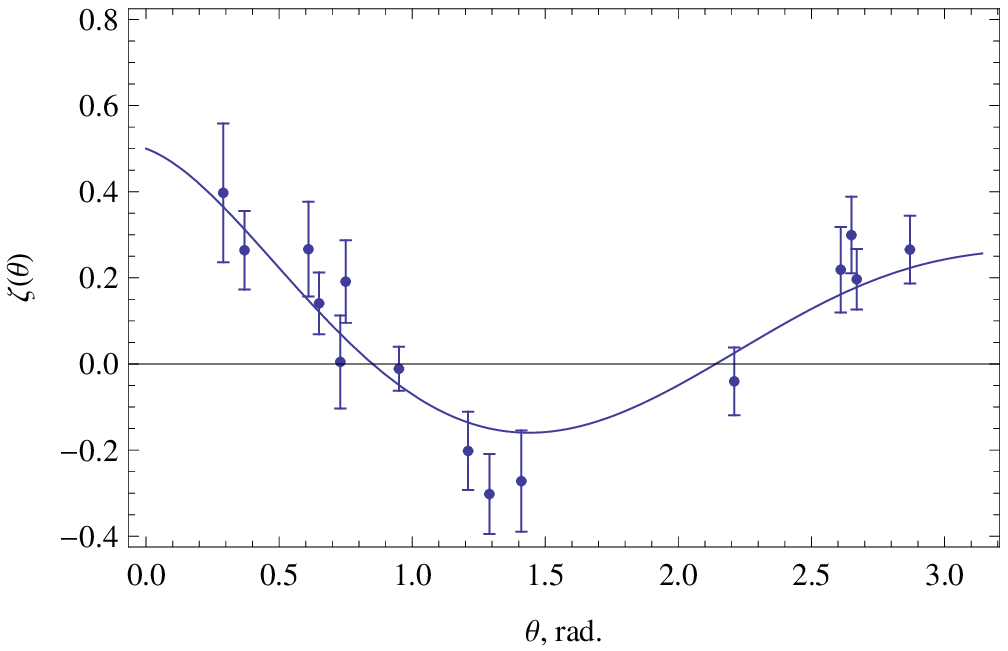}}
\end{minipage}
\hfill
\begin{minipage}{0.5\linewidth}
\vbox{\hbox{\hspace*{4.4cm}b)}\includegraphics[scale=0.72]{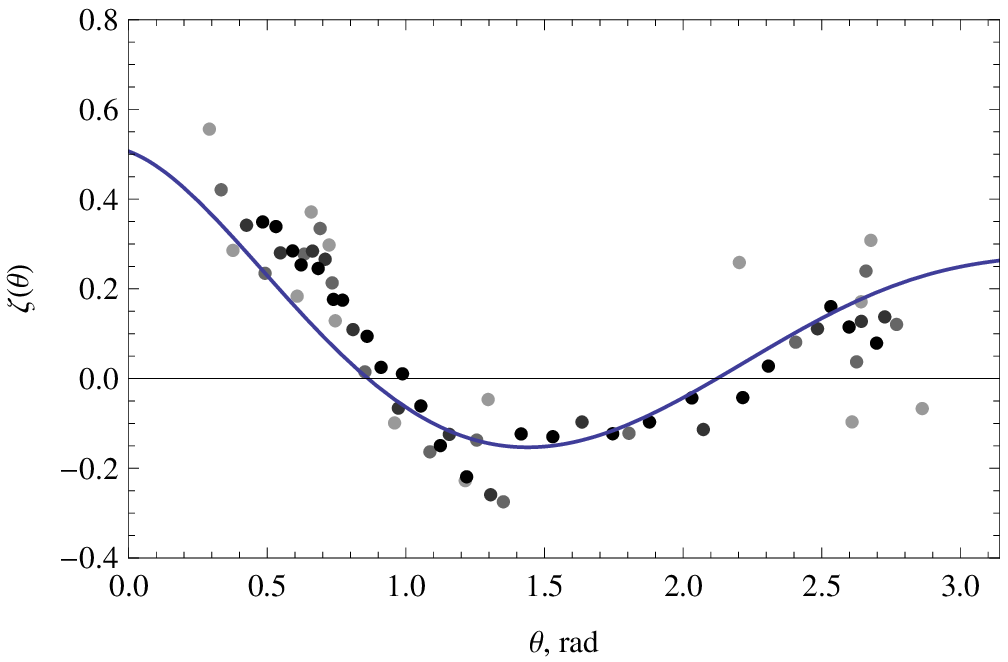}}
\end{minipage}
\caption{a) Results of the numerical simulation of the two-point correlation function for $M=6$ pulsars and $N=50$ TOAs. Signal-to-noise ratio $SNR\simeq 10$. b) Experimental plot of the two-point correlation function. Moving average by 2-5 points were applied. Correlation coefficient between theoretical curve (solid line) and experimental points $\rho = 0.82\pm0.07$
}
\end{figure}

\section{Results}

For simplification of the computer modelling all pulsar data were binned and averaged in $\Delta t=40$ days intervals. Gaps were filled by the linear interpolation of adjacent values. The common part of the observation interval ($\sim 6$ yr) was taken only. Averaging were calculated at the different intervals $15^d\leq\Delta t\leq 60^d$. The discussed effect reveals itself most clearly at $\Delta t\approx40-50^d$. To make the experimental plot of $\zeta(\theta)$ more dense in $\theta$-domain the moving average by 2-5 points was applied to the plot. Fig.~2b shows experimental plot of $\zeta(\theta)$. Correlation coefficient between theoretical curve and experimental points $\rho = 0.82\pm0.07$. For
this correlation coefficient and the number of points $M(M-1)/2 = 15$, the probability that this correlation occurred by chance is ${\rm Pr}(\rho = 0.82\pm 0.07,\; M = 15) < 10^{-4}$ \cite{jenkinswatts:1969}.

\section{Conclusion}

A new approach to the problem of GW detection with pulsar timing array is proposed. This approach is based on  the fact that real physical time series always contain a superposition of noise signals of different kind (with different spectral indices). As a rule the signals have a characteristic amplitude and begin to dominate at characteristic intervals. This feature allows to expand the time series into components of different kind using an appropriate mathematical method. As such a method the singular spectrum analysis was used. Combination of two methods, SSA and pair angular correlation, allowed to detect in TOA residuals a signal (correlated amplitude $\simeq 0.5$ mcs) which at statistically significant level produces similar with GWB effect.

\begin{theacknowledgments}
Author expresses their thanks to all pulsar community for interests and warm support of this approach expressed during the pulsar conference-2010. This work was supported by grant RFBR No.09-02-00584-a.
\end{theacknowledgments}

\bibliographystyle{aipproc}   % if natbib is available

\end{document}